\documentclass[%
 reprint,
 amsmath,amssymb,
 aps,
 prl,
]{revtex4-1}

\usepackage{graphicx}
\usepackage{dcolumn}
\usepackage{bm}
\usepackage[caption=false]{subfig}
\usepackage{natbib}
\usepackage{pdfpages}
\usepackage{pgffor}
\usepackage{hyperref}

\usepackage{color}

\makeatletter
\AtBeginDocument{\let\LS@rot\@undefined}
\makeatother

\begin{document}


\title{Scaling Trapped Ion Quantum Computers Using Fast Gates and Microtraps}

\author{Alexander K. Ratcliffe}
\affiliation{Department of Quantum Science, RSPE, Australian National University}
\email{u6141396@anu.edu.au}
\author{Richard L. Taylor}
\affiliation{Department of Quantum Science, RSPE, Australian National University}
\author{Andr\'e R. R. Carvalho}
\affiliation{Centre for Quantum Dynamics, Griffith University, Gold Coast, QLD 4222, Australia}
\author{Joseph J. Hope}
\affiliation{Department of Quantum Science, RSPE, Australian National University}


\begin{abstract}
	Most attempts to produce a scalable quantum information processing platform based on ion traps have focused on the shuttling of ions in segmented traps. We show that an architecture based on an array of microtraps with fast gates will outperform architectures based on ion shuttling. This system requires higher power lasers, but does not require the manipulation of potentials or shuttling of ions. This improves optical access, reduces the complexity of the trap, and reduces the number of conductive surfaces close to the ions. The use of fast gates also removes limitations on gate time. Error rates of $10^{-5}$ are shown to be possible with $250$mW laser power and a trap separation of $100\mu$m. The performance of the gates is shown to be robust to the limitations in laser repetition rate and the presence of many ions in the trap array.
\end{abstract}

\pacs{03.67.Lx}

\maketitle

The promise of quantum devices to benefit modern computing technology such as the simulation of quantum systems and new encryption technologies, relies on the scaling properties of quantum information processing compared to classical computing \cite{Nielsen2010}. Thus far, no platform has achieved a scale that outperforms classical computers. Architectures based on trapping ions in a single linear trap have achieved all the operations required for viable quantum information processing \cite{Leibfried2003a, Chen2015,Kielpinski2008,Stick2006,Seidelin2006,Harty2014,Leibfried2003a,Monroe1995,Benhelm2008,Debnath2016}, however, like all other platforms, they are currently limited in the scalability of the number of qubits \cite{Wineland1998}.

While many proposed quantum computing platforms can generate large numbers of qubits, the key figure of merit is the number of high-fidelity entangling operations performable over their decoherence time. The speed of the gate operation therefore causes a bound on scalability. Putting more ions in linear Paul traps, for example, requires the trapping frequency to be lowered to prevent buckling of the ion chain \cite{Wineland1998}. This increases the time for sideband-resolving adiabatic gates, which operate slower than the trapping period \cite{Schmidt-Kaler2003}. Attempts to overcome this limitation have focused on several schemes using 2D arrays where ions are moved closer when performing gate operations \cite{Blakestad2009, Hffner2008}, confined in Wigner crystals inside Penning traps \cite{Baltrusch2011FastCrystals,Bohnet2016QuantumIons}, or segmented linear Paul traps, where ions are moved between traps \cite{Kielpinski2002,Amini2010,Reichle2006,Schulz2006,Huber2008,Kaufmann2014}. The fastest demonstrated shuttling processes, which use schemes to relax the adiabatic criteria \cite{Torrontegui2013,Walther2012a}, have been of the order of five trap periods. While it may be possible to push the limits of adiabaticity further, until ions are moved within 1 or 2 trap periods \cite{Blatt2017}, this still limits the number of gate operations achievable before the state decoheres.



Non-adiabatic gates (``fast gates'') were proposed \cite{Garcia-Ripoll2003,Duan2004} to overcome limitations on the gate time posed by sideband-resolving adiabatic gates. Rather than attempting to resolve motional sidebands, fast gate schemes use broadband pulse sequences to entangle the ions, and to restore the ionic motion after the gate operation. Fast gates improve prospects for larger computations in linear Paul traps and have been recently demonstrated experimentally \cite{Schafer2017,Wong-Campos2017,Johnson2017}. Even with fast gates, practical challenges of scalability in these geometries remain \cite{Bentley2015,Bentley2016,Taylor2017}, with computation in a linear Paul trap scaling to 40-50 ions before the error due to compounding gates becomes significant \cite{Taylor2017}. Additionally, addressing individual ions becomes increasingly difficult as their separation decreases \cite{Monroe2013}. 

We propose a method of overcoming these limits using fast gates to produce entangling operations between separate microtraps. An architecture based on trapping ions in individual microtraps has previously been proposed \cite{Cirac2000}, which required a separate ion to be shuttled around the array. A more recent proposal specified a simple electrode design for a series of parallel linear traps, where magnetic field gradients could be used to simulate spin-spin interactions for a quantum simulation \cite{Welzel2011}. This architecture allows for ions to be localised close to the minima of the microtrap potentials, hence resolution of ion location is not limited by the number of ions in the trap. Spatial resolution is then determined by experimental design and the minimum distance is limited by the feature size of the trap.  Fast gates require high laser power, but do not require ion shuttling, magnetic field gradients, or time-dependent potentials. We show that this proposed architecture using fast gates in microtrap arrays provides greater scalability than both current ion shuttling based platforms, and fast gate schemes based on linear traps. We show that the laser parameters required to connect ions between adjoining linear traps with high fidelity are achievable.

\begin{figure}
	\includegraphics[width=0.95 \columnwidth]{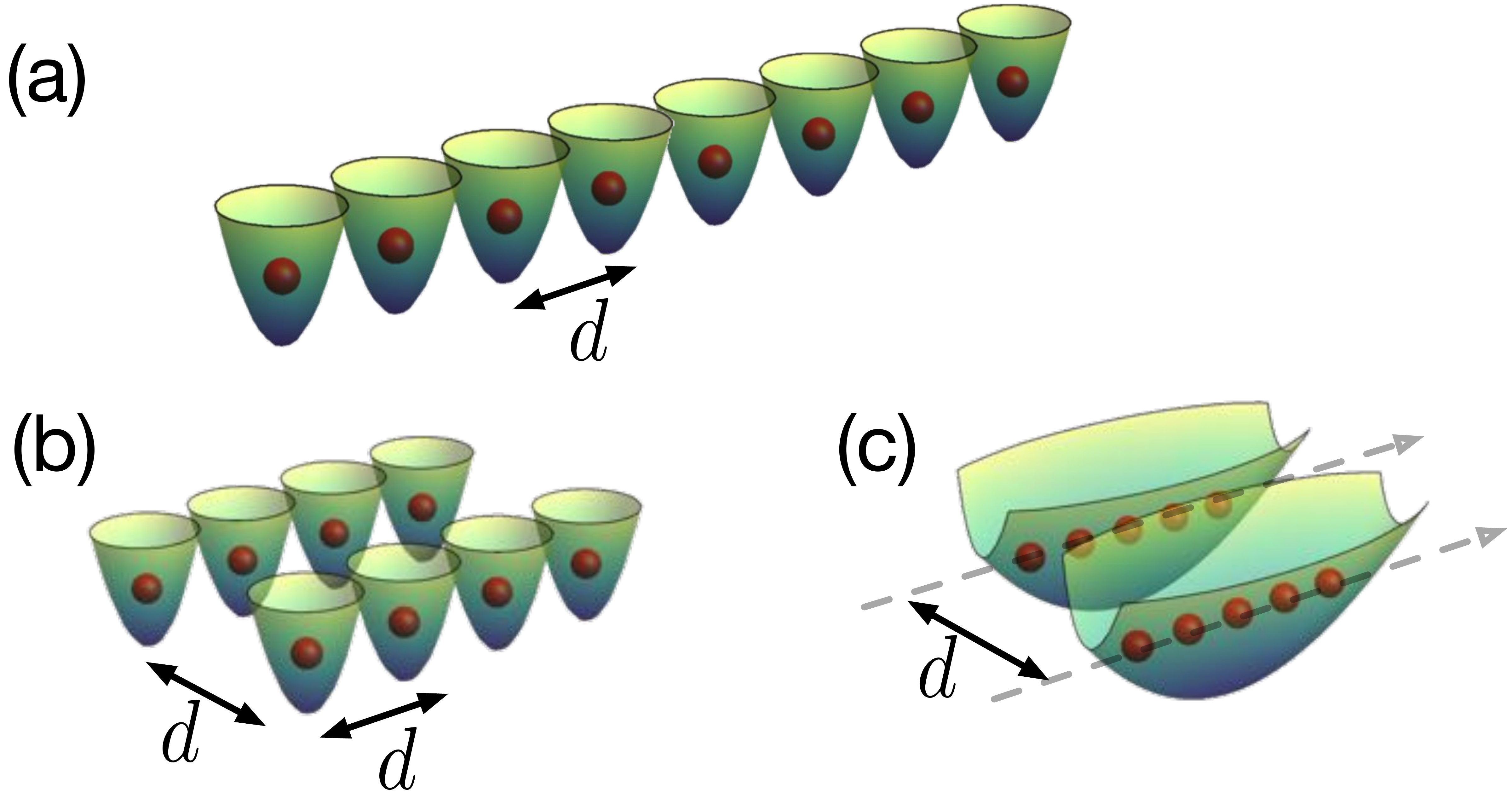}
	\caption{(a) Diagram with a 1D lattice of trapped ions sitting in individual microtraps separated by a distance $d$. In this letter, we model this situation. In the future, more scalable arrangements will likely consist of arrays of traps holding single or multiple ions, as shown in parts (b) and (c). Fast gates can then be used efficiently between nearby ions from different traps, without need for the potentials to be changed, or for the ions to be moved.}
	\label{fig:traps} 
\end{figure}

Multi-qubit gates for trapped ions rely on the strong Coulomb interaction they share. When ions are trapped separately, that Coulomb interaction 
is generally weaker. Rather than attempting to move the ions closer, we find that fast gate schemes can still produce high fidelity gates in less than a trap period. Consider a set of Paul traps each containing a single ion as shown in Fig.~\ref{fig:traps}(a), arranged in a linear chain with the minima between nearest neighbouring traps separated by some distance $d$. Microtrap designs already exist where $d$ is as small as $100\mu$m \cite{Kumph2016, Stick2006}. We examine the feasibility of using fast gates to execute two-qubit gates along this chain, and show that as the length of the ion chain increases, the fidelity does not decay indefinitely. Ultimately, we envisage these gates will operate between neighbouring ions in arrays of individual traps, as shown in Fig.~\ref{fig:traps}(b) and Fig.~\ref{fig:traps}(c). The use of Rydberg ions could also be used to enhance this interaction through their dipole and quadrupole coupling \cite{Higgins2017CoherentIon,Muller2008TrappedGates}.

Fast gates use laser pulses to apply state-dependent momentum kicks on pairs of ions, inducing state-dependent energy shifts through the Coulomb interaction between the ions. Well-chosen strengths and timing of kicks can create a state-dependent phase shift and simultaneously return the motional state of the ions to their initial state. This creates a controlled phase gate $\hat{U}_\text{CPhase}=e^{i \frac{\pi}{4}\sigma^z_1\sigma^z_2}$, which is a sufficient entangling gate for universal computation. Different numbers of pulses and ratios between kick strengths define different schemes. 

The gate schemes we examine here are a generalisation of the Fast Robust Antisymmetric Gate (FRAG) scheme \cite{Bentley2015}, a variant of the GZC scheme \cite{Garcia-Ripoll2003}. They consist of six groups of counter-propagating $\pi$-pulses incident on the ions to be entangled. These pulse groups are defined by fixed ratios of pulses, and some global scaling of pulse number given by the factor $n$. We evaluate gate fidelity as a function of experimental design and required total gate time, and find that the important elements of the experimental design can be reduced to a single dimensionless parameter. Further details of the model, approximations used, and the FRAG scheme can be found in the supplemental material \cite{Rar}.




We use numerical searches to find pulse timings that produce high quality gate operations, with the state-averaged fidelity $F$, given as the fidelity of the post-gate state with the target state integrated over all initial states. This is efficient to compute and strongly related to other distance measures for high-fidelity gates. Within the optimisation, we impose an upper bound on the pulse timings, equivalent to setting a maximum gate time. Optimisation is then run over a set of increasing upper bounds, which allows for a simple numerical optimisation and analysis of the relationship between gate time and infidelity $1-F$. See Section III of the supplemental material for details of the optimisation~\cite{Rar}.

It is sensible report results in terms of $1-F$ because we examine fidelities extremely close to unity. Whilst our numerical calculations use the full infidelity defined above, a Taylor expansion of $1-F$, justified by its small value and given in full in the supplemental material~\cite{Rar}, reveals the important parameters of the problem and helps guide our numerical optimisation.

The first important quantity is the normalised difference between the breathing mode frequency $\omega_{\text{BR}}$, and the common motional mode frequency $\omega$
\begin{eqnarray}
\chi=\frac{\omega_{\text{BR}}-\omega}{\omega},
\label{chi}
\end{eqnarray}
corresponding to the relative spacing of the vibrational mode spectrum. We express this as a function of the experimentally relevant parameter $\xi=\frac{d^3\omega^2}{\alpha}$ with a simpler form, where $\alpha=\frac{e^2}{4\pi\varepsilon_0}\frac{1}{M}$. Here $e$ is the electron charge, $M$ the mass of the ions, and $\varepsilon_0$ the vacuum permittivity. See Section IV of the supplementary material \cite{Rar} for further details of the derivation of $\chi$ and expression in terms of $\xi$.


For a fixed value of the Lamb-Dicke parameter $\eta$, the infidelity is fully described by the number of pulses in each pulse train, and trigonometric expressions of $\chi$.  A derivation of this is shown in the supplementary material \cite{Rar}.  The schemes we consider have either $n$ or $2n$ pulses in each pulse train, where $n$ is a positive integer.  Therefore, $n$ and $\chi$ completely specify the optimal infidelities as a function of the dimensionless gate time $\tau_G$ expressed in trap periods $\tau_G=\frac{\omega~t_G}{2\pi}$.



\begin{figure}
	\subfloat[]{\includegraphics[scale=0.2]{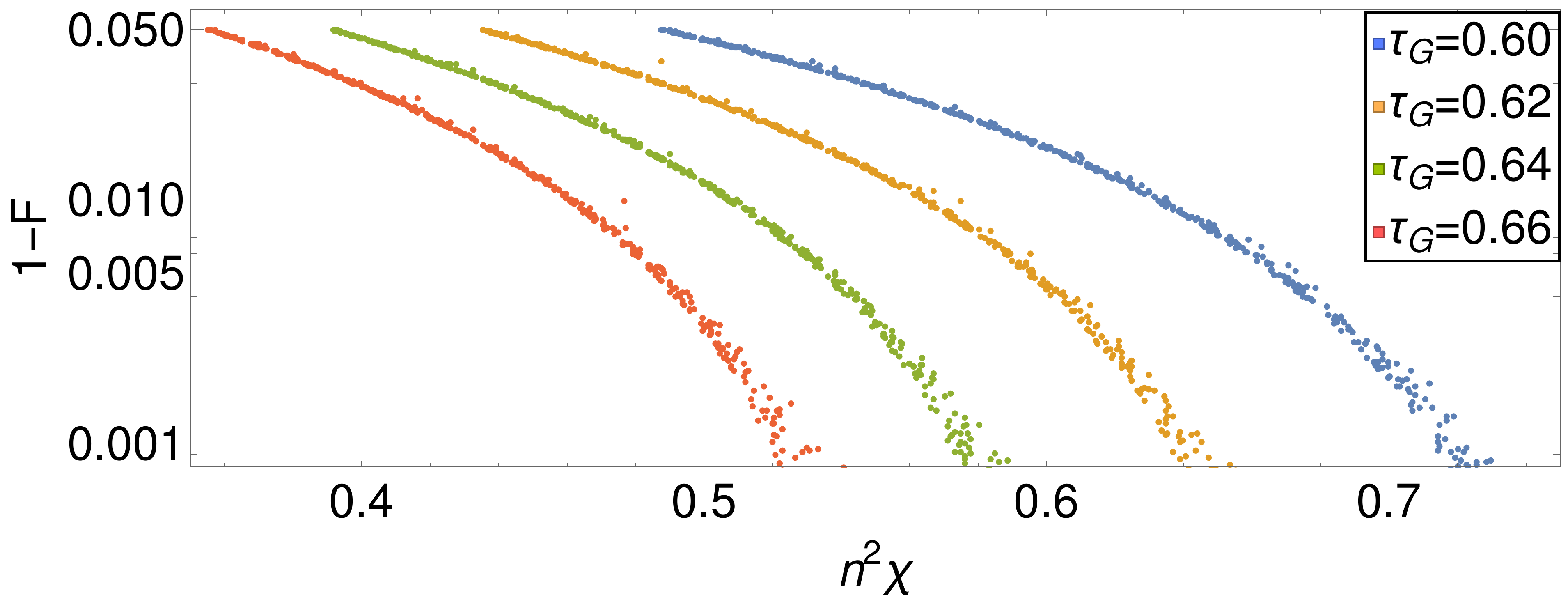}} \\
	\vspace*{-0.3cm} \subfloat[]{ \includegraphics[scale=0.18]{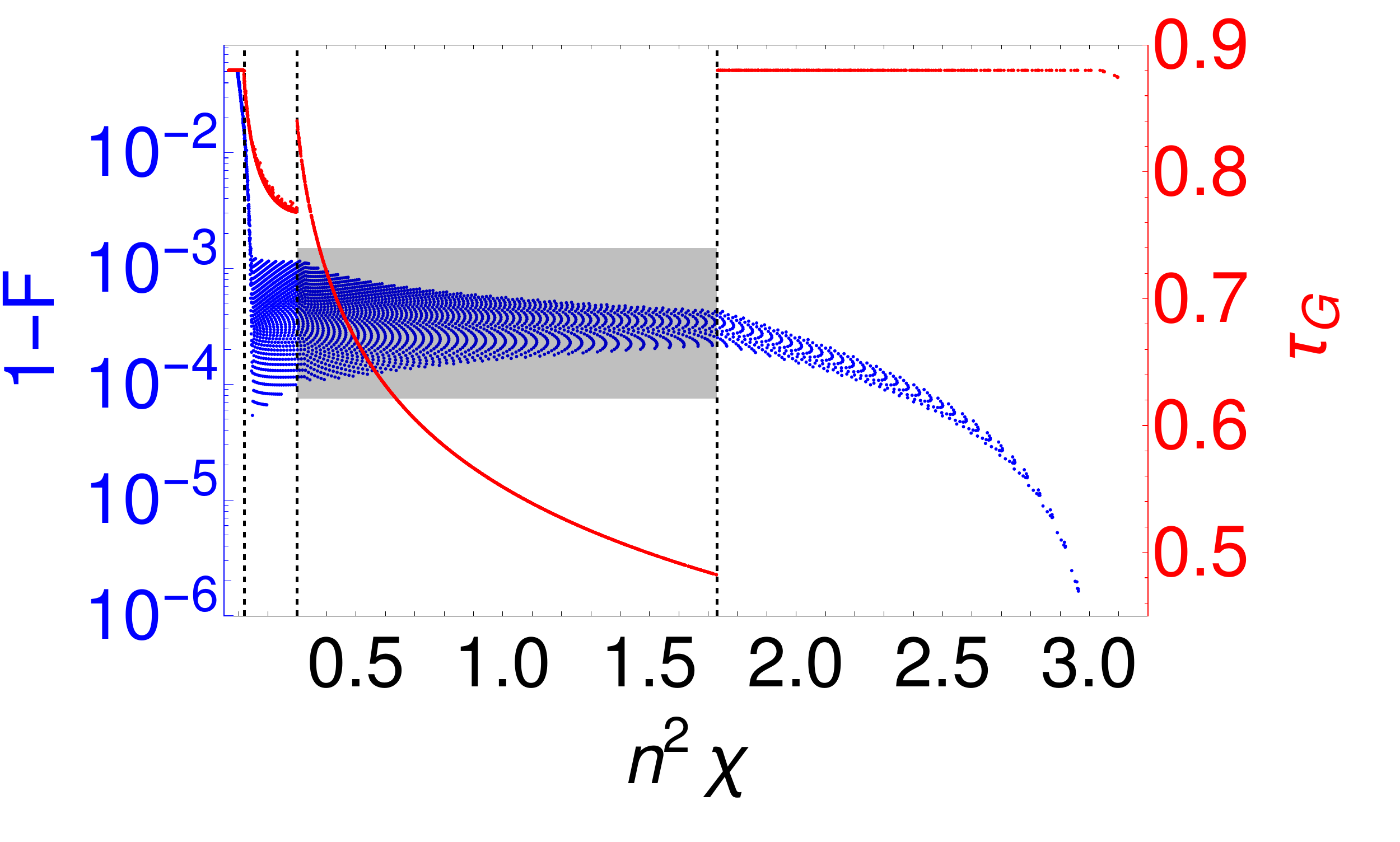}}
	\subfloat[]{ \includegraphics[scale=0.12]{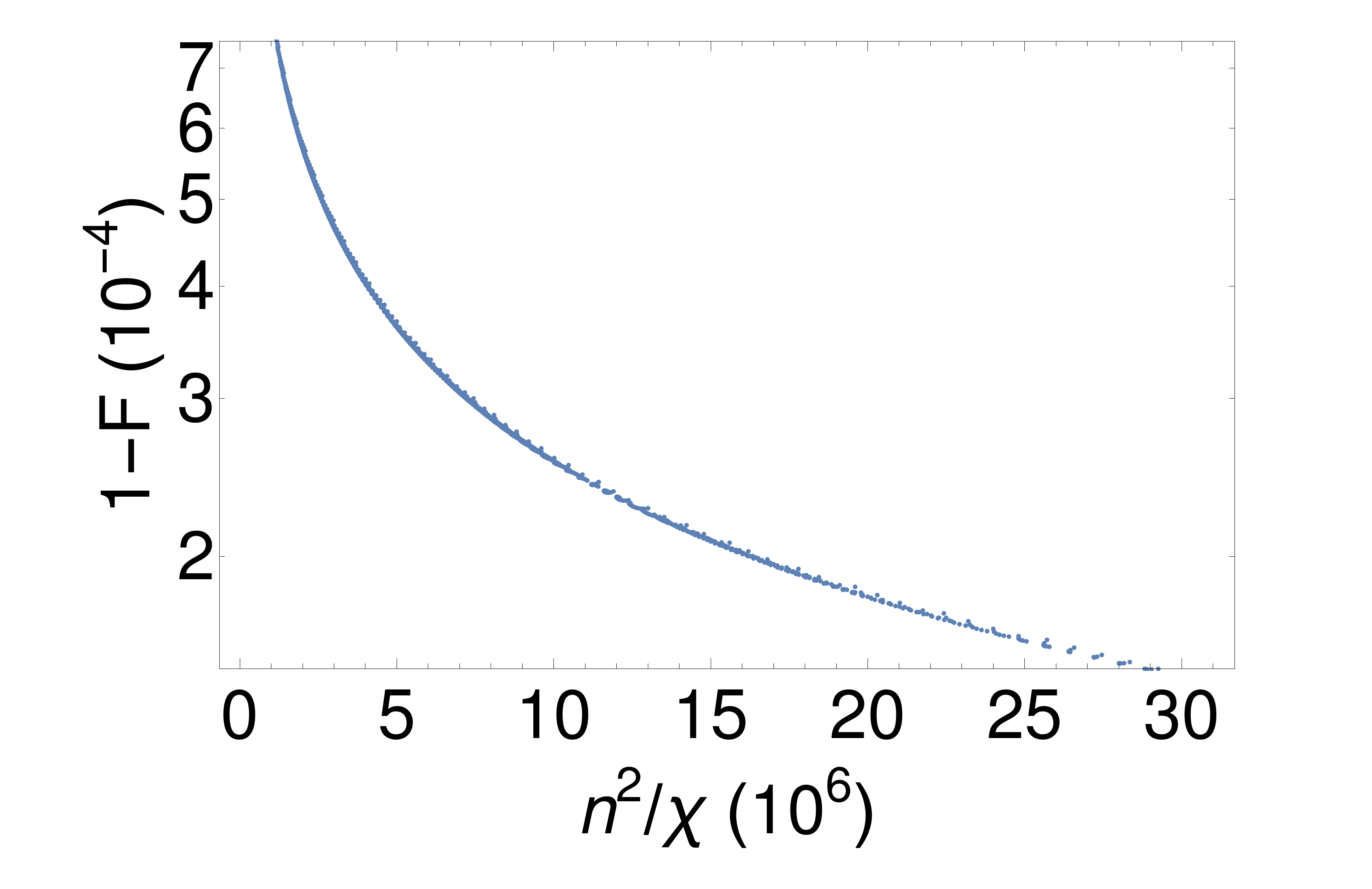} }
	\caption{\label{singleParam} The infidelity of a 2-ion gate plotted for many different values of $n$ and $\chi$. Despite considerable structure in these solutions, there are two types of regions in parameter space where the gate solutions are locally simple. (a) Infidelity as a function of $n^2\chi$ for several gate time caps imposed by the optimisation. This shows that $n^2\chi$ is a useful single dimensionless parameter for predicting infidelity in this parameter region. (b) Infidelity (blue - left vertical axis) and total gate time (red - right vertical axis) as a function of the single non-dimensional parameter $n^2\chi$. We see the optimised solution transition from taking the full allowed time to taking less time than allowed, while the infidelity stops being well defined by $n^2\chi$. For larger values of $n$, new gate solutions of the first type appear, but with much lower infidelities. (c) Infidelity of the gates shown in the grey shaded section of (b), showing that in this region $n^2/\chi$ is a good predictor of infidelity.}
\end{figure}

To analyse the performance of fast gates in microtraps, we find optimised gates for a large range of values of $n$ and $\chi$, as a function of operational gate time $\tau_G$.  For some parameter regimes, the optimised fidelity gate takes the maximum gate time allowed by the optimisation, but for other parameter regimes the highest quality solutions are faster. When our optimised solutions take as long as the gate time upper bound, their infidelities bunch along almost monotonic curves when plotted as a function of $n^2\chi$, shown in Fig.~\ref{singleParam}(a). When they take less time than the optimisation upper bound, shown in Fig.~\ref{singleParam}(b), their infidelities are an almost monotonic function of $n^2/\chi$, as shown in Fig.~\ref{singleParam}(c). As $n$ increases, the achievable infidelity goes down without limit as new classes of solution are found, requiring increasing laser power. We find that with a trap separation of $100\mu$m, a $1$MHz trap and a gate time just under the trap period, a laser power of around $100$mW allows infidelities of $10^{-2}$, which is sufficient to implement fault-tolerant computation using surface codes \cite{Fowler2012SurfaceComputation}. Achieving $10^{-5}$ infidelity to implement earlier proposed schemes \cite{Steane1999EfficientComputing} would require a laser power of $250$mW.

\begin{figure}
	\includegraphics[scale=0.12]{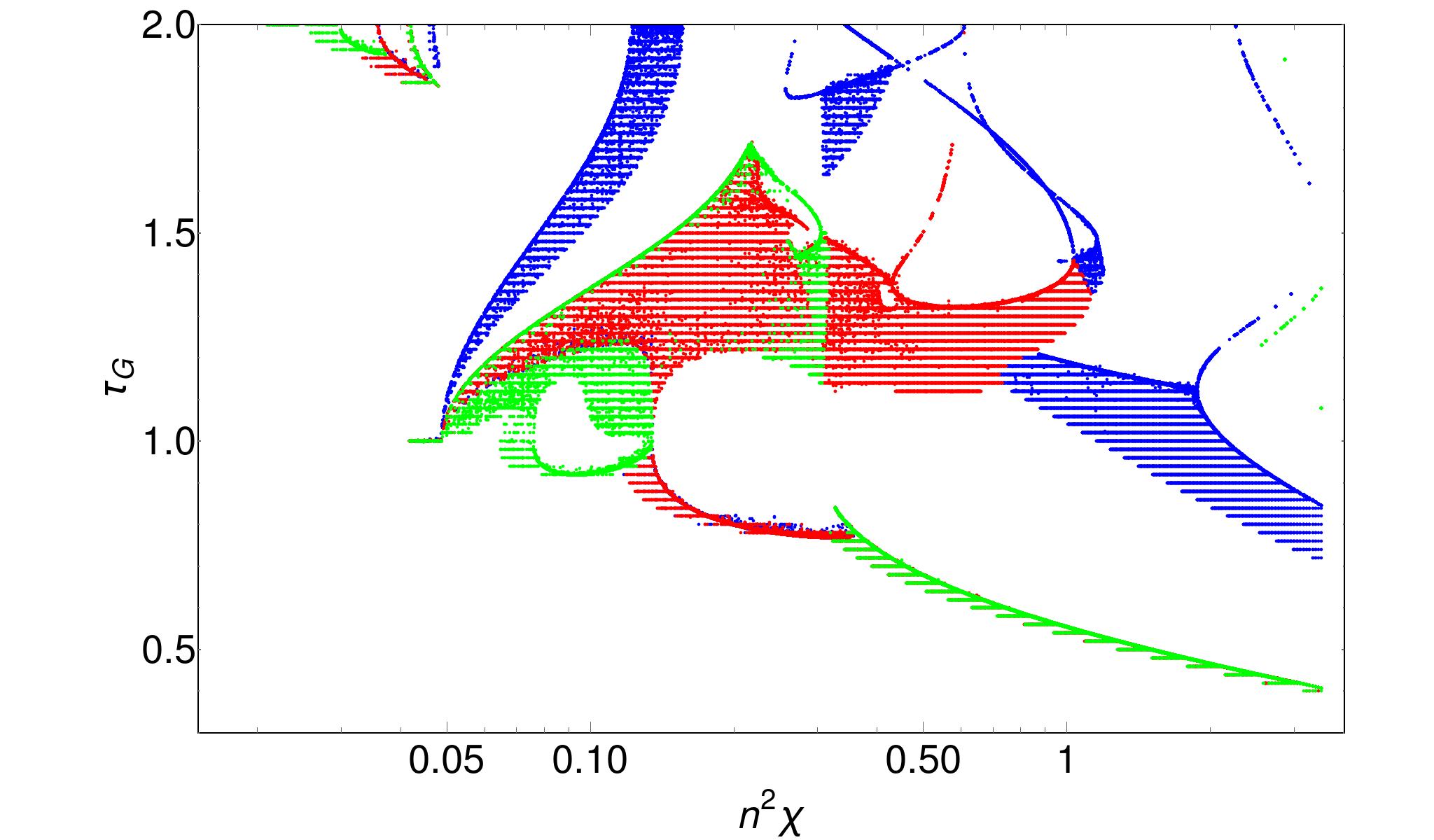}
	\caption{\label{paramSpace} Gate solutions with infidelities less than $10^{-2}$ as a function of gate time in trap periods $\tau_G$ and the parameter $n^2\chi$.  Different orderings of pulses and hence different schemes are denoted by different colours. Here the FRAG scheme is indicated by blue, while the red and green represent schemes with different pulse orderings to the FRAG scheme. This shows that the optimal scheme is generally dependent on the choice of gate time and experimental parameter choice. The empty regions show that sometimes increasing the allowed gate time will not result in decreased gate infidelity unless a threshold gate time is reached, as indicated by the next coloured region.}
\end{figure}

Rather than fixing the maximum gate time and examining the infidelity, we can do the reverse. The distribution of optimised gates with a fidelity greater than $99\%$ is shown in Fig. \ref{paramSpace}. The empty regions in this figure indicate parameter choices where the optimised gate requires less time than the maximum allowed, so there is a gap until a higher-fidelity solution exists with a longer gate time. The different schemes caused by reordering of the kick timings are shown as different coloured regions within Fig.~\ref{paramSpace}. We can see that there are distinct regions where one of these schemes provides the optimal gate, and that the distinct `jumps' in parameter space are associated with a changing scheme. The changes of behaviour in Fig.~\ref{singleParam}(b) correspond to reaching the edge of an empty region in Fig.~\ref{paramSpace}.

The optimal system for gate performance in both infidelity and non-dimensional time will be one that maximises the value of $n$. Depending on the experimental limitations, it may be advantageous to maximise $\chi$; achieved by decreasing the separation of microtraps or decreasing the trapping frequency. However, decreasing the trapping frequency will also increase the total dimensional gate time. It is thus optimal to only decrease the trap separation to improve gate times. The description of performance using the parameter $n^2\chi$ demonstrates that the parameter $n$ has a more significant impact on performance than $\chi$. This indicates that a key focus for improving gate performance should be to increase the number of pulses in a train, and hence the repetition rate of the laser used. 

Thus far, this analysis comes from simulations of pairs of qubits, however the performance for larger quantities of qubits must be examined. Importantly, we must assess the scaling performance as a function of the parameter $\chi$. Maximising $n$ is always beneficial in both time and fidelity, but changing $\chi$ and the number of microtraps both affect the dynamics of the ion chain. We use gates optimized for a simple two-ion system and apply these to a system with more ions. This method will not in general produce the highest fidelity gates for systems with more ions, but it allows for computationally feasible searches. As a result, the gate time and pulse timings remain fixed over a changing number of ions. The fidelity includes the motional states of all the ions in the chain, and the electronic states of the ions not being operated on are implicitly preserved. Hence this provides the complete fidelity for the multi-ion system under a gate operation.

\begin{figure}
	\subfloat[]{ \includegraphics[scale=0.115]{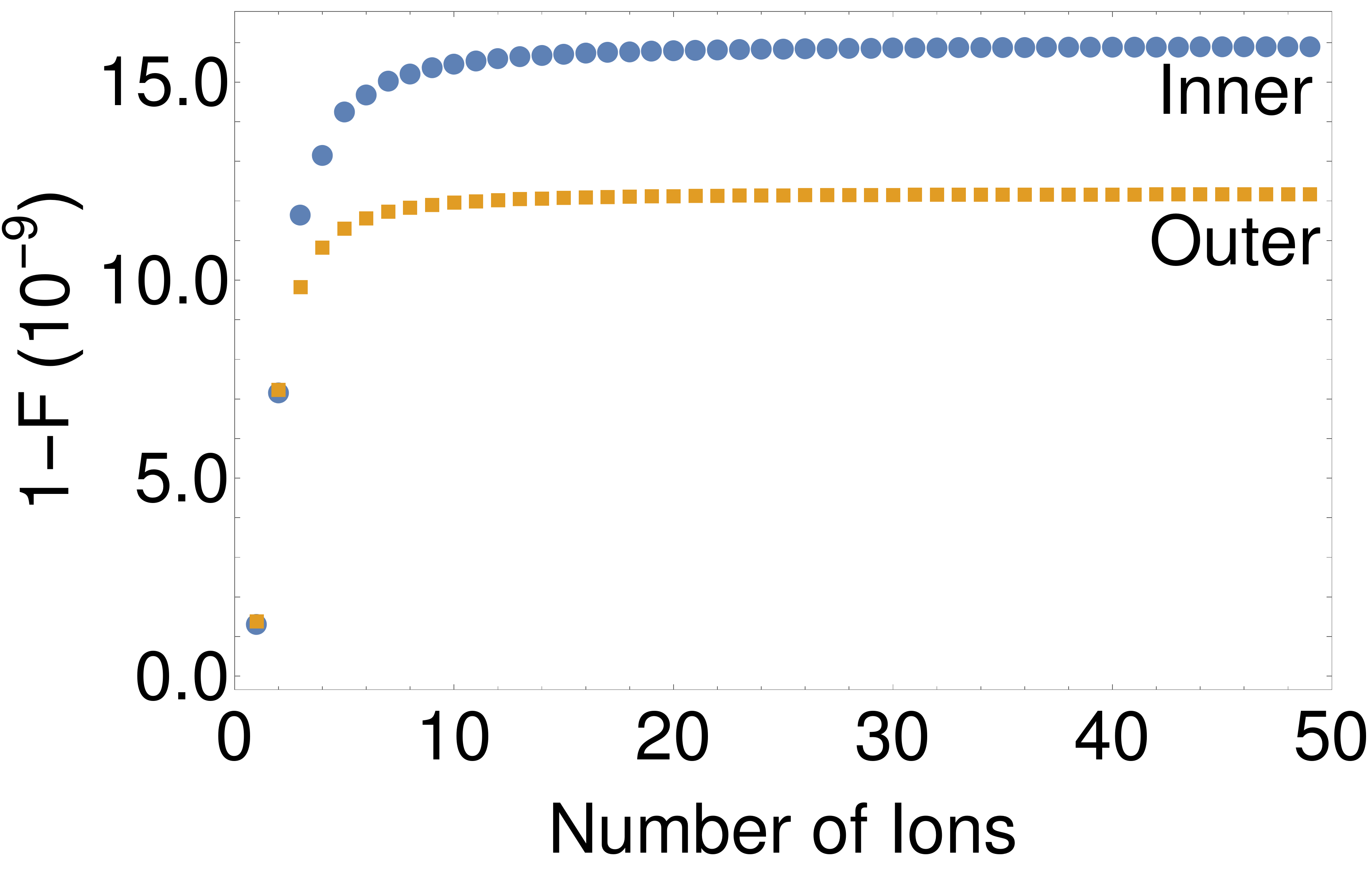} } 
	\subfloat[]{ \includegraphics[scale=0.11]{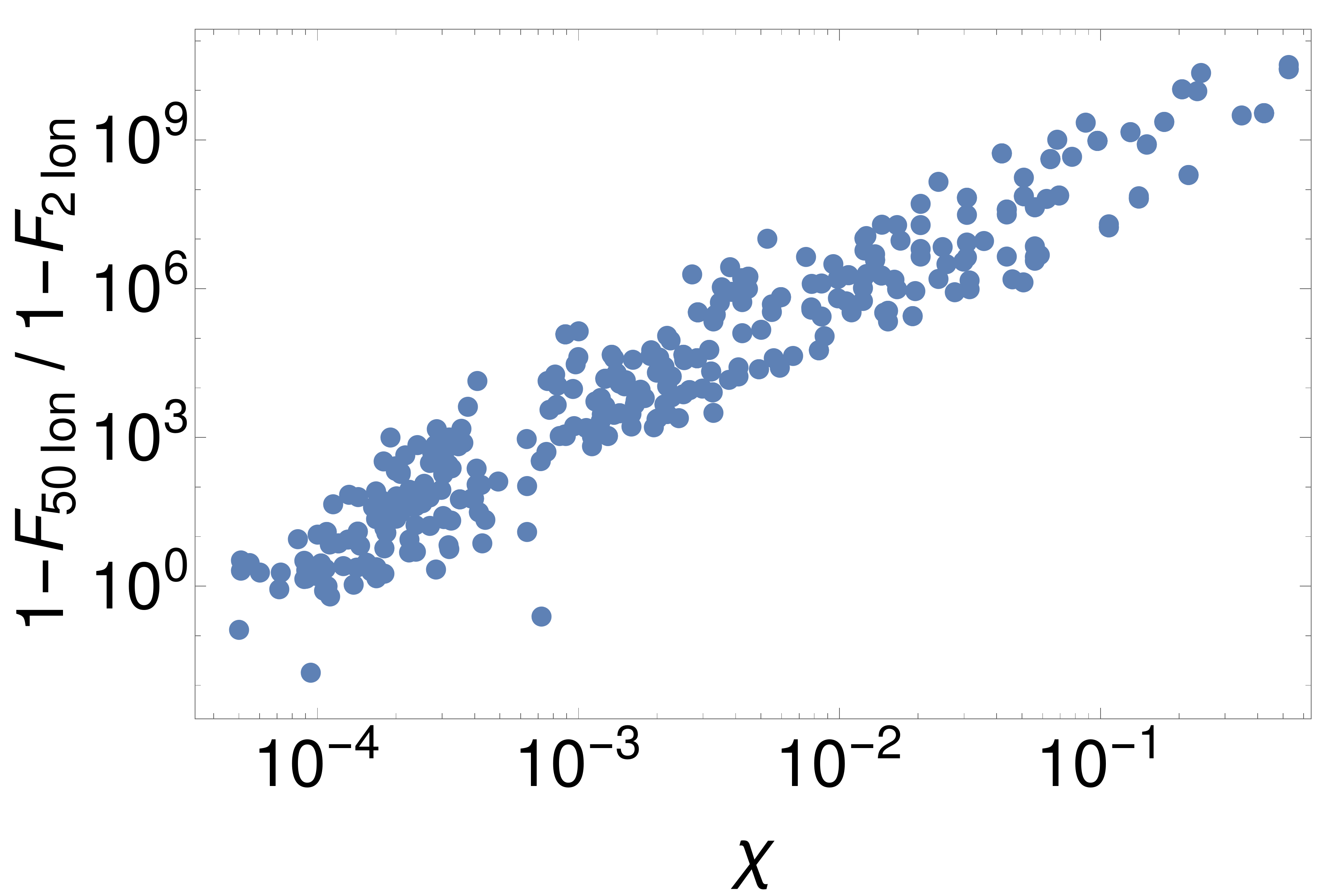} } 
	\caption{\label{ionNum} (a) Infidelity of a 2-ion optimised gate ($n=50$, $\chi=1.8\times10^{-4}$, gate time of $1.4$ trap periods), with fixed pulse timings and gate time, with an increasing number of microtraps in the processor. Shown for both the innermost pair of ions and the outermost pair of ions. Here the infidelity is for the motional and electronic state of all the ions in the chain. (b) Ratios of the infidelity of a gate applied to a 50-ion array to the infidelity of that gate applied to a two-ion array as a function of $\chi$. Showing microtraps with low $\chi$ scale more effectively than linear traps when using gates optimised for two-qubit systems. }
\end{figure}

Fig.~\ref{ionNum}(a) shows that gate fidelities initially decrease as the number of ions is increased, but plateaus for chains of around 10 ions or more. This is due to the decreasing impact of additional ions to the motional modes of the system. This trend is observed when applying gates between any two adjacent ions in the system. Here we have investigated the impact on gates applied to a pair of ions in the middle and on the edge of the chain, as they exhibit the lowest and highest infidelities, respectively.

Trap geometries also affect scaling of two-qubit optimised gates, Fig.~\ref{ionNum}(b) shows that there is a clear dependence on $\chi$. The data was created using a set of randomly chosen high fidelity gates with various values of $\chi$, which were then applied to both a two-ion system and a 50-ion system. The ratio of the infidelities for these two systems shows a clear linear trend in a log-log plot. Here this indicates a cubic relationship between ion scaling performance and the parameter $\chi$. 

\begin{figure}
	\subfloat[]{ \includegraphics[scale=0.115]{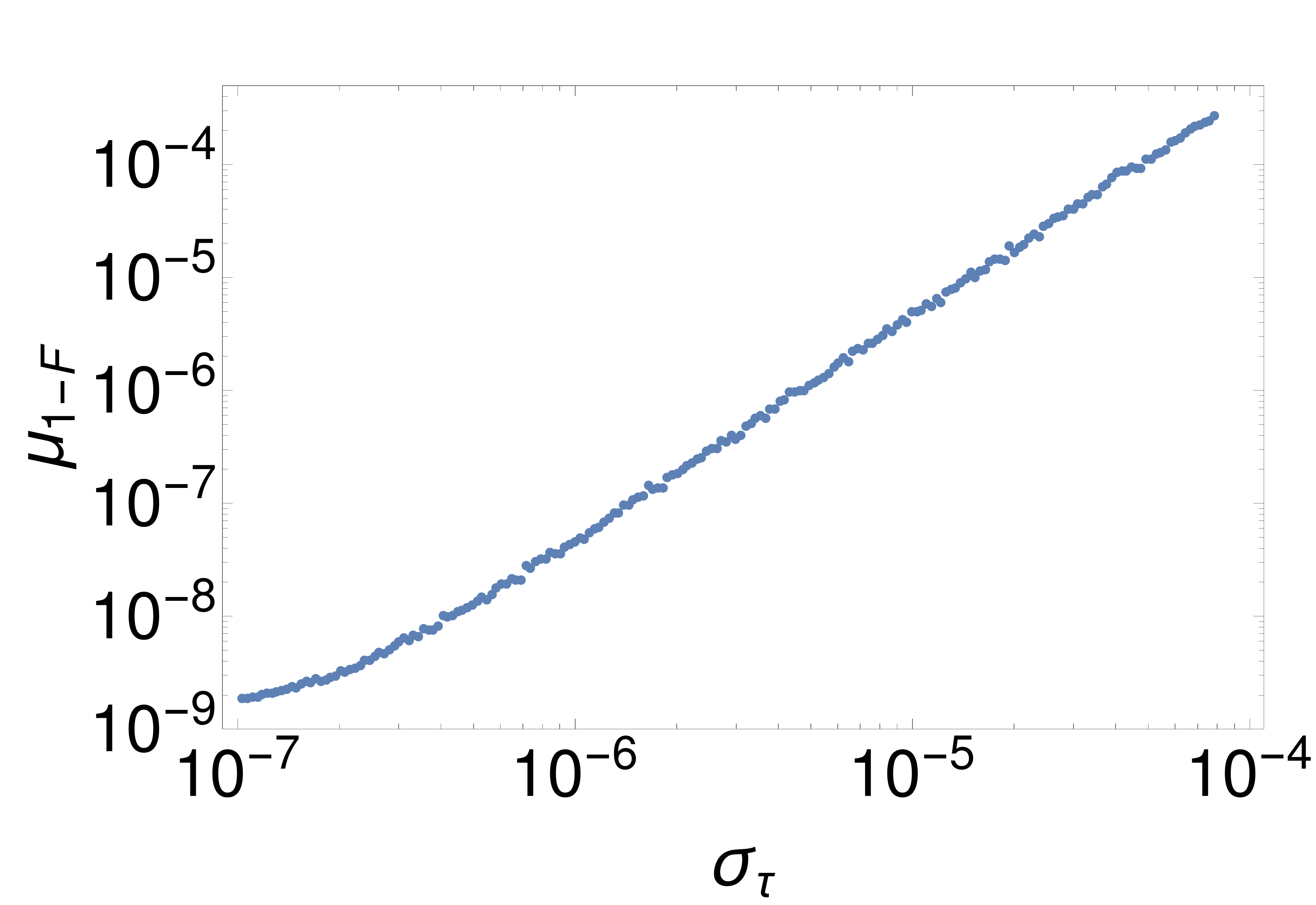} } 
	\subfloat[]{  \includegraphics[scale=0.115]{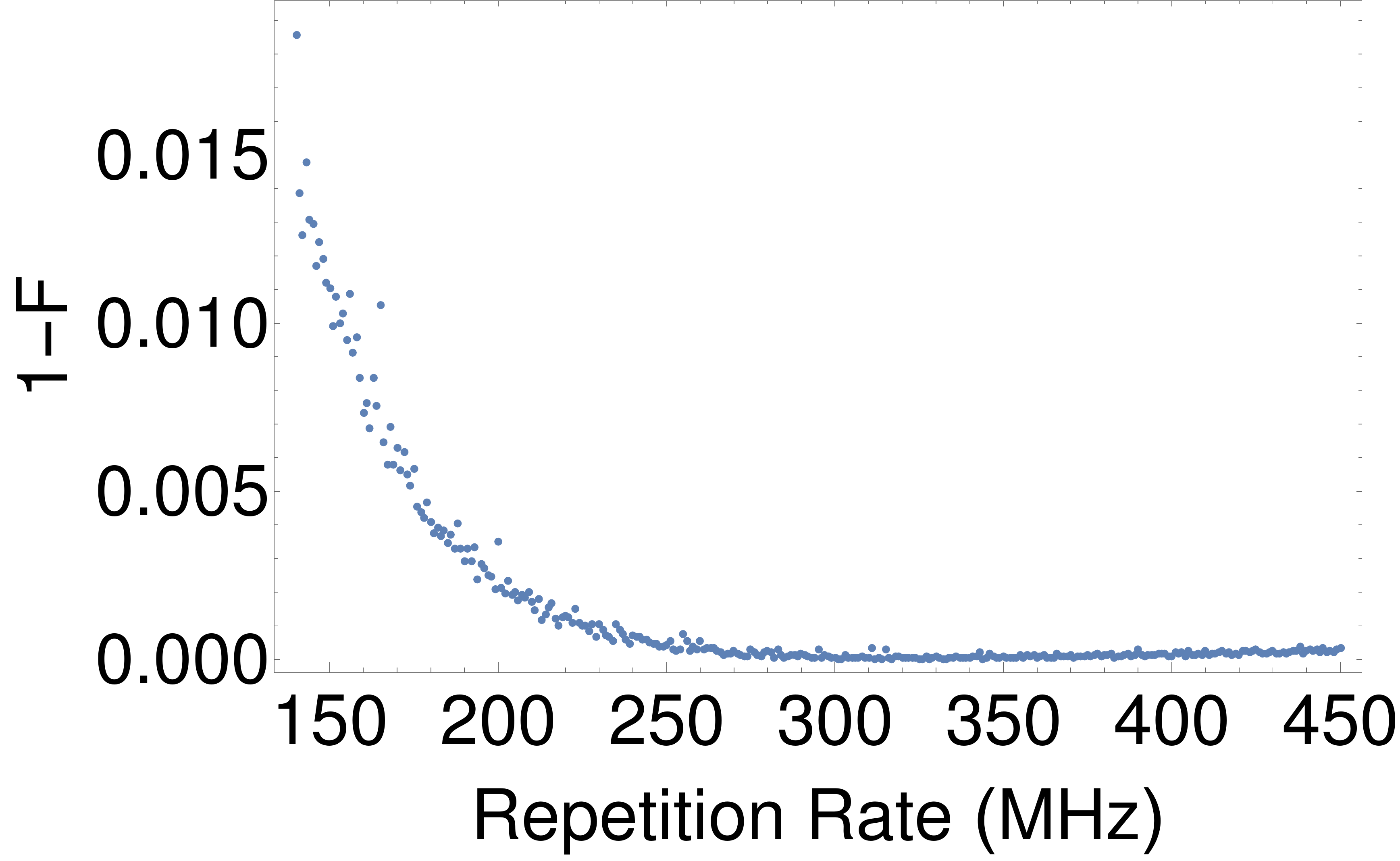} }
	\caption{\label{robust} (a) Mean infidelities of an optimised gate ($n=50$, gate time of $1.4$ trap periods) against the standard deviation of the Gaussian noise applied to the timings, given in absolute trap periods. Demonstrating fidelity is significantly decreased when imperfect pulse timings are applied, this result was consistent for different parameters. (b) An optimised gate ($n=24$, $\omega=2\pi$MHz, gate time of $2.5$$\mu$s), showing the infidelity for varying finite repetition rates. The the minimum repetition rate required to resolve pulse trains is $135$MHz. 
	}
\end{figure}

An important feature of any quantum information processing architecture is the performance of the architecture under realistic noise and experimental error. The first error we investigate is encountered in systems using delay loops to achieve multiple coincident pulses, where there is some error on the timing of the pulses. This error was simulated by applying a Gaussian noise with a varying standard deviation to the gate timings. The resulting distribution of infidelities was observed to be exponential, we use the mean of this distribution as a measure of the average infidelity. The impact of this form of noise on the infidelity is significant, see Fig.~\ref{robust}(a), indicating that heavy attention should be paid to ensure correct timings of gate pulses when using a method of delay loops. Accuracies required for high fidelities are a factor $10^4$ shorter than the trapping period.

The alternate option of using a pulse-picker to generate pulse trains results in both a spreading of the pulses that make up the impulses $n_j$ defined in the FRAG scheme, as well as a small shift in the mean timings $\tau_j$ of those impulses due to the discrete pulse times. We see that the gate infidelity remains remarkably robust to a finite repetition rate. In Fig.~\ref{robust}(b), we see that gates maintain a high fidelity up to the point where different groups of pulses would be required to overlap.
This agrees with results previously obtained for the FRAG gate scheme in linear Paul traps \cite{Bentley2015}. The non-monotonicty of the infidelity is a result of aforementioned shift in the mean gate times. 

Taking all these limitations into account, we conclude that high fidelity fast gates can be executed between microtraps using currently available technology. As an example, we find a scheme that is capable of producing a controlled phase gate in 2.5 trap periods with a fidelity of 99.8\% and 99.995\%, using repetition rates of 200 and 300 times the trapping frequency, respectively. Using $\text{Ca}^\text{+}$ ions, an {\raise.17ex\hbox{$\scriptstyle\sim$}}$1$MHz trapping frequency and 100$\mu$m trap separation ($\chi=1.8\times10^{-4}$), as used in some current experimental set-ups \cite{Kumph2016,Niedermayr2014}, corresponding to a counter-propagating $\pi$-pulse repetition rate of $200$MHz and $300$MHz. This would result in gate times of 2.5$\mu$s. These requirements are consistent with the $300$MHz repetition rates, achieved with a laser power of $190$mW \cite{Hussain2016}. There is a linear relationship between $\pi$-pulse repetition rate and laser power, hence for a $200$MHz repetition rate a laser power of $130$mW should be expected. This is a comparable gate time and infidelity to that reported for experimental fast gates in a linear trap \cite{Schafer2017}, but works across separate microtraps, providing significant benefits to scalability.

In conclusion, we propose that a microtrap architecture with fast gates provides an experimentally realisable platform for dramatically improved scaling for trapped ion QIP platforms. This architecture compares favourably to current state of the art platforms based on ion shuttling, whilst requiring less complexity in the trap geometry. 


It was shown that gate fidelity was surprisingly robust to effects of the finite laser repetition rate for the counter-propagating $\pi$-pulses. The repetition rate needs only be sufficient to maintain separation of pulse trains, which was shown to be experimentally feasible under a standard set of experimental parameters.

It was shown that the fidelities of the fast gates do not decay indefinitely as the number of ions increases. Microtrap architectures also allow multidimensional arrays.  This has significant potential performance benefits, as the penalties for scaling in each dimension are independent, allowing considerably better fidelities for a given number of qubits.  Furthermore, the increased connectivity between distant ions when using nearest-neighbour interactions helps 2D or 3D systems to require many fewer gates to perform any given algorithm.

\begin{acknowledgments}
	The authors thank R. Blatt for providing useful comments and providing details on current ion trapping experiments. This work was supported by the Australian Government through the Australian Research Council's Discovery Projects funding scheme (project DP130101613). 
\end{acknowledgments}

\bibliographystyle{bibsty}

\widetext

\section*{}


\section*{Supplementary material}

# Supplementary Material

## I. MODEL

The model used is given as a set of individual traps (microtraps) arranged in a single line. Each microtrap is simply a Paul trap that contains only a single ion, it is thus generally not elongated along an axis as linear Paul traps are. The potential energy in this situation is then given as the sum of the trappinging potentials and the Coulomb potentials between the ions, as given in Eq. 1.

$$V = \frac{e^2}{4\pi\varepsilon_0} \sum_{i=1}^{N-1} \sum_{j=i+1}^{N} \frac{1}{((j-i)d + x_j - x_i)} + \frac{1}{2}M\omega^2 \sum_{i=1}^{N} x_i^2 \tag{1}$$

where $x_i$ is the position of the $i^{\text{th}}$ ion in the chain relative to the centre of its own trap, $d$ is the separation between each microtrap, $\omega$ is the angular trapping frequency of each microtrap, $M$ is the ion mass and $N$ the number of ions in the chain.

For efficient simulation of the motion of ions, we use a normal mode expansion. This approximates the motion in terms of $N$ oscillatory modes, each mode described by some frequency of oscillation $\omega_p$ and coupling to the ions $\vec{b}_p$. This is done by linearising the potential around the ions stationary points and is valid for sufficiently small displacements of the ions around their stationary points. The motion of the $i^{\text{th}}$ ion is then given by

$$x_i = \sum_{p=1}^{N} A_p b_p^i \sin(\omega_p t + \phi_p), \tag{2}$$

with the amplitude of each mode $A_p$ and the phase of each mode $\phi_p$ determined by initial conditions.

The error associated with this approximation can be estimated by the upper limit of the size of the next order term in the expansion at maximum displacement, which is of the order of $10^{-5}$ for the parameters used in this analysis ($\omega = 2\pi \times 10^6$ Hz, $d = 10^{-4}$ m). This approximation could not be made for significantly lower trapping frequencies and smaller ion spacings.

## II. GATE SCHEME

We analyse fast gate schemes that use a series of broadband counter-propagating $\pi$-pulses, incident on the two ions to which the gate is to be applied. These $\pi$-pulses can be simple square pulses of the appropriate height, or shaped for convenience of production or robustness of the change of state. They are always used in counter-propagating pairs so that they do not change the internal state of the ions, but give them a state-dependent momentum kick. These kicks are significant due to the Lamb-Dicke parameter, which is typically of the order of $\eta \approx 0.16$. These state-dependent kicks then have a state-dependent effect on the energy (and phase evolution) of those modes through interaction via the Coulomb potential. Correctly chosen kicks can ultimately return the motional state of the ions to their initial state, leaving the net effect of a controlled phase gate:

$$\hat{U}_{\text{CPhase}} = e^{i\frac{\pi}{4}\sigma_1^z \sigma_2^z} \tag{3}$$

A general fast gate can be described by a set of pulse timings $\vec{t}$ and pulse-group intensities $\vec{z}$. Different pulse-group intensities are generated by having different numbers of single pulses comprising a pulse-group. These single pulses then arrive at, or symmetrically around, the pulse time given for that group, hence the pulse intensities are given as as integer multiples. Multiple fast gate schemes have been proposed in the literature, such as: GZC [1], Duan [2] and FRAG [3]. Each scheme imposes a different set of restrictions on the number of distinct pulses, symmetry and different ratios of pulse numbers in pulse groups. For the FRAG scheme, the timings of theses pulses and the number of pulses in each pulse group are given by the vectors $\underline{t}$ and $\underline{z}$ respectively:

$$\begin{aligned} \underline{t} &= (-\tau_1, -\tau_2, -\tau_3, \tau_3, \tau_2, \tau_1), \\ \underline{z} &= (-n, 2n, -2n, 2n, -2n, n). \end{aligned} \tag{4}$$

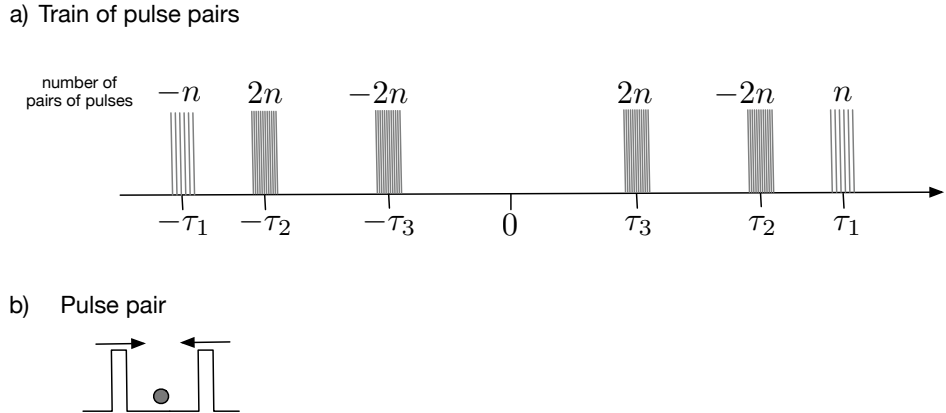

FIG. 1. a) Diagram with the pulse timing for the FRAG scheme. The components $z_j$ of the $z$ vector indicate the number of pairs of pulses that hit the ion at each time $\tau_j$. The sign in $z_j$ indicates which pulse within each pair (shown in b) reaches the ion first. This gives the sign of the momentum kick imprinted on the ion.

The sign of the components of $\underline{z}$ corresponds to changing the direction of the initially incident pulse, and the factor of $n$ is an integer that characterizes the overall scale of numbers of pulses at each time. With an infinite repetition rate laser, To produce an effective C-Phase gate the timings $(\tau_1, \tau_2, \tau_3)$ are chosen to give the desired gate. In the original FRAG scheme proposal there was a strict ordering on the magnitude of $(\tau_1, \tau_2, \tau_3)$. In this implementation we do not impose a strict ordering of the times $(\tau_1, \tau_2, \tau_3)$, effectively resulting in a set of six possible pulse schemes. The total gate time is therefore twice the maximum of the values of $\tau_1$, $\tau_2$, and $\tau_3$.

### III. OPTIMISATION

We use numerical searches to find pulse timings that produce high quality gate operations, with the state-averaged fidelity $F$, given as the integral of the square of the norm of the overlap between the post-gate state with the target state integrated over all initial states. This is efficient to compute and it is strongly related to other distance measures for high-fidelity gates. As we examine fidelities extremely close to unity, we report the infidelity $1 - F$. This is a function of the phase mismatch $\Delta\phi$ around the target $\pi/4$ phase, and the population changes of the motional modes, as given by:

$$\Delta\phi = |\sum_p 8\eta^2 \frac{\omega}{\omega_p} b_p^1 b_p^2 \sum_{i \neq j} z_i z_j \sin(\omega_n |t_i - t_j|) - \frac{\pi}{4}|$$

and

$$\Delta P_p = 4\eta \sqrt{\frac{\omega}{\omega_p}} \sum_k z_k \sin(\omega_p t_k)$$

where $\Delta P_p$ is the population changes of the $p$-th motional mode.

For efficient computation of two-ion gates, we further simplify this measure by using a truncated expansion of the infidelity in these variables:

$$1 - F \approx \frac{2}{3}(\Delta\phi)^2 + 0.8(\Delta P_1)^2 + 0.8(\Delta P_2)^2. \tag{5}$$

While this approximate form is efficient for generating gate schemes, we use the full form when reporting achievable fidelities, for example, in the presence of multiple ions.

We can see from Eq. (5) that the infidelity for a two-ion system, $1 - F$, depends on the Lamb-Dicke parameter $\eta$, the angular frequencies collective motional modes $\omega_n$, the coupling of the $k$-th ion to the $p$-th mode, $b_p^k$, and the number of pulses in the $i$-th pulse train $z_i$. The collective mode frequencies $\omega_n$ can be calculated from the mass of the ions $M$, the separation of the microtraps $d$, and the trapping frequency $\omega$ of the individual microtraps.

We search for pulse timings that produce optimal gate fidelity within a given time bound. This optimisation is run as a set of local gradient searches in the three-dimensional parameter space of the pulse timings, over a large set

of initial gate sequences. The highest fidelity of these local optimisations is then taken to be the optimal gate for that cap in the gate time. Note that the optimal gate occasionally takes less time than the maximum allowed. By increasing the cap in total gate time and repeating this process, we map out the optimal fidelity for fast gates as a function of gate time.

## IV. DIMENSIONLESS PARAMETER

We see from Eq. 5 that the system behaviour depends on the ratios of the frequencies of the collective modes. These are in turn functions of the geometry, and the dimensionless parameter $\xi = \frac{d^3 \omega^2}{\alpha}$, where $\alpha = \frac{e^2}{4\pi\varepsilon_0} \frac{1}{M}$. Here $e$ is the electron charge, $M$ the mass of the ions, and $\varepsilon_0$ the vacuum permittivity. For a two-ion system, there is only one ratio, so it entirely characterises the behaviour.

We define $\chi$ as the normalised difference between the breathing mode frequency and the common motional mode frequency $\chi = \frac{\omega_{\mathrm{BR}} - \omega}{\omega}$ which can be expressed in terms of the more fundamental parameter $\xi$ as given in Eq. 6.

$$\chi = \sqrt{\frac{1}{3}(9 - \beta\gamma^{\frac{1}{3}} + \beta\gamma^{\frac{2}{3}})} - 1 \qquad (6)$$

where

$$\gamma = 1 + \frac{3(9 + \sqrt{3}\sqrt{27 + 2\xi})}{\xi}$$

and

$$\beta = 9 - \sqrt{3}\sqrt{27 + 2\xi}$$

Even for three or more ions, the system is still well characterised by $\chi$, which is the normalised gap to the lowest energy excitation in the system. This is because it defines the rate of relative acquisition of phase between the excited and unexcited modes. Its value lies in the range between 0 and $\sqrt{3} - 1$. The upper bound corresponds to the limit where both microtraps are merged, which is the case for standard linear trap geometries.

We also observe a factor of $n^2$ appearing in all fidelity terms, as $z_i$ scales with $n$. Assuming a fixed value for $\eta$ we then have that, together, $n$ and $\chi$ give a full description of $\Delta\phi$ and $\Delta P_n$ in conjunction with the gate times. Therefore, they completely specify the optimal infidelities as a function of the dimensionless gate time $\tau_G$ expressed in trap periods $\tau_G = \frac{\omega t_G}{2\pi}$.

---



\end{document}